\documentclass[amssymb, preprintnumbers, showpacs, showkeys,aps,prb,superscriptaddress,twocolumn,nobibnotes]{revtex4-1}

\usepackage{pgfplots}
\usepackage{tikz}
\usetikzlibrary{positioning}

\usepackage{color} 
\usepackage{xcolor} 
\usepackage{hyperref}
\usepackage{slashed}
\usepackage{graphicx}
\usepackage{amsmath}
\usepackage{latexsym}
\usepackage{epstopdf}
\usepackage{amsmath}
\usepackage{enumitem}
\usepackage{amssymb,amsmath}
\usepackage{multirow}
\usepackage{mathtools}
\usepackage{bbm}
\usepackage{bm}
\usepackage{amsfonts}
\usepackage{amssymb}
\usepackage{calligra}
\usepackage{calrsfs}
\usepackage{makecell}
\usepackage[normalem]{ulem}
\usepackage[USenglish,american]{babel}

\expandafter\ifx\csname package@font\endcsname\relax\else
 \expandafter\expandafter
 \expandafter\usepackage
 \expandafter\expandafter
 \expandafter{\csname package@font\endcsname}%
\fi
\begin{document}

\title{Nonlinear wave propagation in large extra spatial dimensions and the blackbody thermal laws}

\author{I. Soares}%
\email{winacio@cbpf.br}
\affiliation{Centro Brasileiro de Pesquisas Físicas, Rua Dr. Xavier Sigaud, 150, URCA, Rio de Janeiro CEP 22290-180, RJ, Brazil}%

\author{R. Turcati}%
\email{turcati@cbpf.br}
\affiliation{Centro Brasileiro de Pesquisas Físicas, Rua Dr. Xavier Sigaud, 150, URCA, Rio de Janeiro CEP 22290-180, RJ, Brazil}%

\author{S. B. Duarte}%
\email{sbd@cbpf.br}
\affiliation{Centro Brasileiro de Pesquisas Físicas, Rua Dr. Xavier Sigaud, 150, URCA, Rio de Janeiro CEP 22290-180, RJ, Brazil}%

\begin{abstract}

Nonlinear wave propagation in large extra spatial dimensions (on and above $d=2$) is investigated in the context of nonlinear electrodynamics theories that depend exclusively on the invariant $\mathcal{F}\left(=-(1/4)F_{\mu\nu}F^{\mu\nu}\right)$. In this vein, we consider propagating waves under the influence of external uniform electric and magnetic fields. Features related to the blackbody radiation in the presence of a background constant electric field such as the generalization of the spectral energy density distribution and the Stefan-Boltzmann law are obtained. Interestingly enough, anisotropic contributions to the frequency spectrum appear in connection to the nonlinearity of the electromagnetic field. In addition, the long wavelength regime and Wien's displacement law in this situation are studied. The corresponding thermodynamics quantities at thermal equilibrium, such as energy, pressure, entropy, and heat capacity densities are contemplated as well. 

\end{abstract}

\maketitle

\section{Introduction}\label{introduction}

In the past decades, there has been an emergence of physical models involving spacetimes with extra dimensions to solve open problems in modern physics\cite{Biesiada:2001iy,Oda:2001ss,Bander:2001qk}. The main effort in these theories includes scenarios involving the unification of the fundamental interactions, such as the Kaluza-Klein\cite{Coquereaux:1988ne} and superstring theories\cite{Green:1987sp,Green:1987mn}. The dimensionality also plays an important role in other areas of physics, such as path integrals formulation\cite{Fukutaka:1986ps,Neves:2000hw}, phase space and momentum operators\cite{Bashir:2001ad}, topological defects \cite{Jiang:1999xh}, among others\cite{Arkani-Hamed:1998lzu,DiFrancesco:2001xur,Ito:2001fd,Ivashchuk:1999rm}.

Introducing extra dimensions to the structure of spacetime has the potential to modify the fundamental laws as we know\cite{Al-Jaber2003,Cardoso:2005cd,Alnes:2005ed}. In this sense, it would be expected small deviations from the well-known established results in physics. Theories such as Kaluza-Klein and string theories propose compactified extra dimensions, which could induce deviations from the standard results at high energies. On the other hand, large infinitely extra dimensions, while a nonphysical situation, can be very useful, from the theoretical perspective, to get an insight about the fundamental physical laws\cite{Alnes:2005ed}. 

The purpose of the present paper is twofold. First, we study wave propagation in arbitrary infinitely large spatial dimensions from the perspective of nonlinear theories. In this vein, it will be considered theories that depend exclusively on the invariant $\mathcal{F}$. Second, features related to blackbody radiation in the presence of an external electric field and the consequences to the thermodynamics quantities in this context will be contemplated. 

The remainder of this paper is organized as follows. In Sec. (\ref{generalframework}) we review the main features of gauge and Poincarè invariant nonlinear electrodynamics theories in large extra spatial dimensions, while the wave propagation in background electromagnetic fields is derived in Sec. (\ref{MDR}). Aspects related to the blackbody spectral density, generalized Stefan-Boltzmann law, long wavelength regime, and Wien's law are discussed in Secs. (\ref{spectralradiance}), (\ref{sectiongeneralizedSB}) and (\ref{longwave}), respectively. The implications on the thermodynamics quantities such as energy, pressure, entropy, and specific heat densities are contemplated in Sec. (\ref{thermoproperties}). Our final remarks and conclusions can be found in Secs. (\ref{further}) and (\ref{conclusions}).

In what follows we will work in Gaussian units. In our conventions, the signature of the Minkowski metric is $\left(+,-,-,-\right)$.


\section{The nonlinear wave propagation in extra spatial dimensions}\label{generalframework}

In this section, we will derive the nonlinear wave propagation in spacetimes with additional large infinitely spatial dimensions. Such a description can be achieved when one considers the wave propagation in an external electromagnetic field as a weak field disturbance propagating around a background classical field. 

To begin with, we consider the general class of density lagrangians in the $(d+1)-$dimensional Minkowski spacetime $M^{d+1}$, i.e., $d-$spatial dimensions and one-time dimension, which is given by 
\begin{eqnarray}\label{lagrangian}
\mathcal{L}=\mathcal{L}\left(\mathcal{F}\right),    
\end{eqnarray}
where 
\begin{eqnarray}\label{bilinear1}
\mathcal{F}&\equiv&-\frac{1}{4}F_{\mu\nu}F^{\mu\nu},
%
\end{eqnarray}
is a invariant bilinear form, $F_{\mu\nu}\left(\equiv\partial_{\mu}A_{\nu}-\partial_{\nu}A_{\mu}\right)$ is the field-strength of the electromagnetic field and $A^{\mu}$ is the associated gauge field. In $M^{d+1}$, the gauge field $A^{\mu}$ has $(d+1)-$components. Concerning the spacetime dimensions, it is well-known that there are no magnetic fields in $d=1$, which prevents the existence of electromagnetic waves\cite{Cardoso:2005cd}. Therefore, we are restricted to considering spatial dimensions $d>1$ to have wave propagation. We also would like to stress the fact that the dual tensor $\tilde{F}^{\mu\nu}$, which is related to the Levi-Civita symbol, will not be considered here because it depends on the spacetime dimension. Since our formalism intends to incorporate a general result for arbitrary spacetime dimensions, the dual of the Maxwell tensor ${F}^{\mu\nu}$ will then be neglected.

The corresponding field equations for the system can be obtained through the Euler-Lagrange equations, namely,
\begin{eqnarray}\label{fieldequation}
\partial_{\nu}\left(\frac{\partial\mathcal{L}}{\partial{F_{\mu\nu}}}\right)=0.
\end{eqnarray}

The full description of the nonlinear system also includes the Bianchi identity, 
\begin{eqnarray}\label{Bianchi}
\partial_{\alpha}F_{\mu\nu}+\partial_{\mu}F_{\nu\alpha}+\partial_{\nu}F_{\alpha\mu}=0.
\end{eqnarray}

Next, by adopting the linearization procedure, the electromagnetic field $F^{\mu\nu}$ can be split into a
background field $F^{\mu\nu}_{B}$ plus a propagating wave $f^{\mu\nu}$ as \begin{equation}\label{split}
F^{\mu\nu}=F^{\mu\nu}_{B}+f^{\mu\nu}.
\end{equation}

Introducing relation (\ref{split}) into the Eq. (\ref{fieldequation}), and assuming that the background field satisfies the Euler-Lagrange equations, one finds 
\begin{eqnarray}\label{linearizedequation}
\partial_{\nu}\left(\Omega^{\mu\nu\alpha\beta}f_{\alpha\beta}\right)=0, 
\end{eqnarray}
where
\begin{eqnarray}
\Omega^{\mu\nu\alpha\beta}=\frac{\partial^{2}\mathcal{L}}{\partial{F_{\mu\nu}}\partial{F_{\alpha\beta}}}\bigg|_{B}. 
\end{eqnarray}

We remark that the tensor $\Omega^{\mu\nu\alpha\beta}$ is symmetric concerning the exchange of the pairs of indices $\mu\nu$ and $\alpha\beta$, and antisymmetric concerning the exchange of indices within each pair. 

If one prefers, it is possible to work at the level of the linearized density lagrangian, which takes the form
\begin{eqnarray}
\mathcal{L}=\frac{1}{2}f_{\mu\nu}\Omega^{\mu\nu\alpha\beta}f_{\alpha\beta}. 
\end{eqnarray}

From the density lagrangian (\ref{lagrangian}), the field equations associated to the wave field $f_{\mu\nu}$ are \begin{eqnarray}
c_{1}\partial_{\nu}f^{\mu\nu}-\frac{1}{2}M^{\mu\nu\alpha\beta}_{B}\partial_{\nu}f_{\alpha\beta}=0, 
\end{eqnarray}
where
\begin{eqnarray}
M^{\mu\nu\alpha\beta}&=&d_{1}F^{\mu\nu}F^{\alpha\beta}, 
\end{eqnarray}
and
\begin{eqnarray}\label{coefficients}
c_{1}&=&\frac{\partial\mathcal{L}}{\partial\mathcal{F}}\bigg|_{\mathbf{E,B}}, \quad 
%
%
d_{1}=\frac{\partial^{2}\mathcal{L}}{\partial\mathcal{F}^{2}}\bigg|_{\mathbf{E,B}}.
\end{eqnarray}

The tensor $M^{\mu\nu\alpha\beta}$ has the same set of symmetries as $\Omega^{\mu\nu\alpha\beta}$. In addition, the coefficients $c_{1}$ and $d_{1}$ are all evaluated at the external fields and carry the nonlinearity of the electromagnetic field.

By considering the regime of slowly varying background electromagnetic fields, Eq. (\ref{linearizedequation}) in Fourier space, i.e., using the plane waves ansatz $a^{\mu}\approx\tilde{a}^{\mu}e^{ik\cdot{x}}$, assumes the following form:
\begin{eqnarray}\label{omega}
\Omega^{\mu\nu\alpha\beta}k_{\nu}f_{\alpha\beta}=0, 
\end{eqnarray}
where $k^{\mu}$ is the wave four-vector associated with the plane wave direction.

Because of the Bianchi identity, the wave field $f^{\mu\nu}$ can be written as 
\begin{eqnarray}\label{bianchigauge}
f_{\mu\nu}=\partial_{\mu}a_{\nu}-\partial_{\nu}a_{\mu}, 
\end{eqnarray}
where $a^{\mu}$ is the gauge field.

Inserting (\ref{bianchigauge}) into (\ref{omega}), one promptly finds 
\begin{eqnarray}
\Omega^{\mu\nu\alpha\beta}k_{\nu}k_{\beta}a_{\alpha}=0,
\end{eqnarray}
where 
\begin{eqnarray}\label{tensor}
\Omega^{\mu\nu\alpha\beta}=c_{1}\left(\eta^{\mu\alpha}\eta^{\nu\beta}-\eta^{\mu\beta}\eta^{\nu\alpha}\right)-M^{\mu\nu\alpha\beta}_{B}.
\end{eqnarray}
For simplicity, we have neglected the tilde-symbol in $\tilde{a}$. 

The above tensor contains an isotropic part plus an anisotropic contribution $M^{\mu\nu\alpha\beta}_{B}$, which is related to the nonlinear behavior of the electromagnetic field in $M^{d+1}$.

\subsection{Dispersion relations in the presence of external electromagnetic fields}\label{MDR}

Let us start off our considerations by deriving the wave modes for the electromagnetic field. For this purpose, one needs to solve the system of linear equations
\begin{eqnarray}
A^{\mu\alpha}\epsilon_{\alpha}=0, 
\end{eqnarray}
where 
\begin{eqnarray}
A^{\mu\alpha}=\Omega^{\mu\nu\alpha\beta}k_{\nu}k_{\beta}, 
\end{eqnarray}
and $\epsilon_{\mu}=a_{\mu}/\sqrt{a^{2}}$ is the normalized polarization four vector. The $a^{2}=a_{\mu}a^{\mu}$ is the quadratic modulus of the gauge field {$a^{\mu}$} associated to the wave field $f^{\mu\nu}$. The four-vector $\epsilon_{\mu}$ denotes the polarization states of the electromagnetic wave.

According to relation (\ref{tensor}), we have
\begin{eqnarray}\label{generaloperator}
A^{\mu\alpha}\equiv{c}_{1}\left(\eta^{\mu\alpha}k^{2}-k^{\mu}k^{\alpha}\right)-M^{\mu\nu\alpha\beta}k_{\nu}k_{\beta}.  
\end{eqnarray}

Since the theory under consideration is gauge invariant, we need to fix the gauge. For convenience, we then adopt the temporal gauge $a^{0}=0$, which split the system of linear equations in
\begin{eqnarray}\label{temporalequation}
A^{0i}\epsilon_{i}=0, 
\end{eqnarray}
and the reduced system
\begin{eqnarray}\label{spatialequation}
A^{ij}\epsilon_{j}=0.
\end{eqnarray}

A nontrivial solution of the above system of linear equations can be found if one finds the vanishing determinant of $A^{ij}$, i.e.,
\begin{eqnarray}
det\left(A^{ij}\right)=0.    
\end{eqnarray}

It is important to note that the approach developed here is valid for flat spacetimes in $(d+1)-$dimensions, i.e., for large infinitely extra dimensions.

In the next section, the corresponding wave frequencies for background uniform electromagnetic fields will be derived. To simplify our task, the external uniform magnetic and electric fields will be evaluated separately.

\subsubsection{The purely magnetic field case}

The first case to be considered is for a pure background uniform magnetic field, i.e., $F_{ij}\ne0$. In this case, the electric field $F_{0i}=E_{i}$ is zero, and assuming $k^{\mu}=\left(w/c,\mathbf{k}\right)$, Eq. (\ref{temporalequation}) becomes 
\begin{eqnarray}\label{}
\mathbf{k}\cdot\mathbf{\epsilon}=0, 
\end{eqnarray}
which reduces to the Coulomb gauge in $M^{d+1}$. Therefore, the electromagnetic wave holds $(d-1)-$degrees of freedom. Note that the temporal gauge and the Coulomb gauge form the generalized radiation gauge in large extra dimensions.

With regards to the Eq. (\ref{spatialequation}), one needs to take the pure spatial sector of the operator (\ref{generaloperator}), i.e., the term $A^{ij}$, which gives us the following form:
\begin{eqnarray}
%
A_{ij}=\left(\frac{w^{2}}{c^{2}}-\mathbf{k}^{2}\right)\delta_{ij}+\frac{d_{1}}{c_{1}}F_{il}F_{jk}k_{l}k_{k}.
\end{eqnarray}

Then, computing the determinant of $A^{ij}$, the nontrivial solution provides us the following wave angular frequencies: 
\begin{eqnarray}
w_{1}\left(\mathbf{k}\right)&=&ck\sqrt{1-\frac{d_{1}}{c_{1}}(\overleftrightarrow{F}\cdot\hat{k})^{2}},\\
w_{2}\left(\mathbf{k}\right)&=&ck,\\
\vdots\nonumber\\
w_{d-2}\left(\mathbf{k}\right)&=&ck, 
\end{eqnarray}
where we have defined $(\overleftrightarrow{F}\cdot\hat{k})_{j}\equiv{F}_{ij}k_{j}$. 

From the above relation, we find $(d-1)$ angular wave frequencies, which are related to the physical propagating modes. Furthermore, the influence of the external magnetic field alters only one of the angular frequency modes, leading to the appearance of the birefringence in large extra dimensions. We also remark that $d_{1}\rightarrow0$, or, equivalently, whenever $F_{ij}\rightarrow{0}$, the $w_{1}$ frequency reduces to $w_{1}\left(\mathbf{k}\right)=ck$. To ensure that $w_{1}$ is real and positive-definite, the condition $c_{1}>{d_{1}}(\overleftrightarrow{F}\cdot\hat{k})^{2}$ must be satisfied. 

An interesting case is when $d=3$, and the magnetic field takes the following form $F_{ij}=-\epsilon_{ijk}\phantom{,}B_{k}$ in Euclidean space, where $B_{k}$ is the magnetic field in three spatial dimensions and $\epsilon_{ijk}$ is the spatial part of the Levi-Civita symbol, which yields the corresponding wave angular frequencies associated to the two polarization states:
\begin{eqnarray}
w_{1}\left(\mathbf{k}\right)&=&ck\sqrt{1-\frac{d_{1}}{c_{1}}(\mathbf{\hat{k}}\times\mathbf{B})^{2}},\\
w_{2}\left(\mathbf{k}\right)&=&ck,
\end{eqnarray}
in accordance with previous results\cite{Bialynicka-Birula:1970nlh,Neves:2022jqq,Soares:2023gfe}.

The group velocity $v_{g(i)}=\partial{w}/\partial{k_{i}}$ related to the set of frequencies $(24)-(26)$, in turn, are given by 
\begin{eqnarray}
v_{g}^{(1)}\left(\mathbf{k}\right)&=&c\left[\frac{\hat{k}_{i}-(d_{1}/c_{1})\left(\overleftrightarrow{F}\cdot\hat{k}\right)_{j}F_{ji}}{\sqrt{1-\frac{d_{1}}{c_{1}}\left(\overleftrightarrow{F}\cdot\hat{k}\right)^{2}}}\right],\\
v_{g}^{(2)}\left(\mathbf{k}\right)&=&c\mathbf{\hat{k}},\\
\vdots\nonumber\\
v_{g}^{(d-2)}\left(\mathbf{k}\right)&=&c\mathbf{\hat{k}}. 
\end{eqnarray}

The group velocity vectors have components in the directions of $\hat{k}$ and the external magnetic field $F_{ij}$, which, in $d=3$, takes the following form\cite{Bialynicka-Birula:1970nlh,Neves:2022jqq,Soares:2023gfe} 
\begin{eqnarray}\label{groupvelocity}
\mathbf{v}_{g}^{(1)}&=&c\frac{\left[c_{1}\mathbf{\hat{k}}-d_{1}\mathbf{B}\times\left(\mathbf{\hat{k}}\times\mathbf{B}\right)\right]}{c_{1}\sqrt{1-\frac{d_{1}}{c_{1}}\left(\mathbf{\hat{k}}\times\mathbf{B}\right)^{2}}},\\
v_{g}^{(2)}\left(\mathbf{k}\right)&=&c\mathbf{\hat{k}}.
\end{eqnarray}

Furthermore, in the absence of the external magnetic field, i.e., whenever $F_{ij}\rightarrow0$, then  $v_{g}^{(1)}=c\mathbf{\hat{k}}$. 

To conclude, we would like to stress that the nature of the magnetic field depends on the spacetime dimension. In the $(3+1)-$case, the magnetic field is a vector. On the other hand, in other dimensions, $F_{ij}$ no longer forms a vector. This aspect is a direct consequence of the antisymmetry of the stress tensor $F^{\mu\nu}$. The electric field $F_{0i}$ has $d-$components, which is a vector field for any spacetime dimension, while the $F_{ij}$ has $d(d-1)/2$ components, and forms a vector only in $d=3$.


\subsubsection{The background electric field case}

Now, for the case of a pure background uniform electric field $\mathbf{E}$, Eq. (\ref{temporalequation}) yields 
\begin{eqnarray}
\mathbf{k}\cdot\mathbf{\epsilon}=-\frac{d_{1}}{c_{1}}\left(\mathbf{k}\cdot\mathbf{E}\right)\left(\mathbf{E}\cdot\mathbf{\epsilon}\right), 
\end{eqnarray}
while the reduced system of linear equations Eq. (\ref{spatialequation}) takes the form
\begin{eqnarray}\label{matrixequation}
\left[\left(\frac{w^{2}}{c^{2}}-\mathbf{k}^{2}\right)\delta_{ij}-\frac{d_{1}}{c_{1}}(\mathbf{k}\cdot\mathbf{E})k_{i}E_{j}+\frac{d_{1}}{c_{1}}\frac{w^{2}}{c^{2}}{E}_{i}{E}_{j}\right]\epsilon_{j}=0.\nonumber\\
\end{eqnarray}

An explicit calculation of the determinant of the $A^{ij}$ operator gives us the following frequencies:
\begin{eqnarray}\label{frequencyelectric1}
w_{1}\left(\mathbf{k}\right)&=&ck\sqrt{1-\frac{d_{1}\left(\mathbf{E}^{2}-(\mathbf{\hat{k}}\cdot\mathbf{E})^{2}\right)}{c_{1}+d_{1}\mathbf{E}^{2}}}, \\
w_{2}\left(\mathbf{k}\right)&=&ck,\\
\vdots\nonumber\\
\label{frequencyelectric2}w_{d-2}\left(\mathbf{k}\right)&=&ck. 
\end{eqnarray}

As in the case of the purely magnetic field, the phenomenon of birefringence appears here. Furthermore, there are $(d-1)-$wave propagating modes. The frequency (\ref{frequencyelectric1}) is real and positive definite if the condition $c_{1}+d_{1}(\mathbf{\hat{k}}\cdot\mathbf{E})^{2}>0$ is satisfied. In the particular case $d=3$, the dispersion relation reduces to two frequency modes, which are given by
\begin{eqnarray}\label{frequency1}
w_{1}\left(\mathbf{k}\right)&=&ck\sqrt{1-\frac{d_{1}(\mathbf{\hat{k}}\times\mathbf{E})^{2}}{c_{1}+d_{1}\mathbf{E}^{2}}},\\
w_{2}\left(\mathbf{k}\right)&=&ck,
\end{eqnarray}
which, again, yields the same result as previous works\cite{Bialynicka-Birula:1970nlh}.

The wave frequency (\ref{frequencyelectric1}) can also be written in terms of the angle $\theta_{d-2}$ between the wave vector $\mathbf{k}$ and the external uniform electric field $\mathbf{E}$, which assumes the form
\begin{eqnarray}
w_{1}\left(\mathbf{k}\right)=ck\sqrt{1-\frac{d_{1}E^{2}}{c_{1}+d_{1}E^{2}}sin^{2}\theta_{d-2}}.    
\end{eqnarray}

The group velocity for the frequencies (\ref{frequencyelectric1}) and (\ref{frequencyelectric2}) are, respectively, given by
\begin{eqnarray}\label{groupvelocity1}
v_{g}^{(1)}\left(\mathbf{k}\right)&=&c\left[\frac{c_{1}\hat{k}_{i}+d_{1}(\mathbf{\hat{k}}\cdot\mathbf{E})E_{i}}{\left(c_{1}+d_{1}\mathbf{E}^{2}\right)\sqrt{1-\frac{d_{1}\left(\mathbf{E}^{2}-\left(\mathbf{k}\cdot\mathbf{E}\right)^{2}\right)}{c_{1}+d_{1}\mathbf{E}^{2}}}}\right],\nonumber\\\\
v_{g}^{(2)}\left(\mathbf{k}\right)&=&c\mathbf{\hat{k}},\\
\vdots\nonumber\\
v_{g}^{(d-2)}\left(\mathbf{k}\right)&=&c\mathbf{\hat{k}}. 
\end{eqnarray}

The group velocity $v_{g}^{(1)}$ has components in the direction of $\hat{k}$ and $E_{i}$. In the absence of a background electric field, the group velocity related to the mode $w_{1}(\mathbf{k})$ reduces to $v_{g}=c\mathbf{\hat{k}}$.

\section{The blackbody radiation in extra dimensions}

Now we would like to understand the effects of the nonlinear electromagnetic field in extra dimensions on the photon gas in thermal equilibrium. To accomplish that, we first consider nonlinear electromagnetic waves trapped in a box of volume $V$ in $ d-$ dimensional space in thermal equilibrium at the temperature T with the enclosure. The spin-statistics is assumed to be valid in higher dimensions\cite{Weinberg:1984vb}. Furthermore, each photon of zero rest mass has $d-1$ degrees of freedom. In addition, since the magnetic field depends on the specific spacetime dimension, we will then consider only an external uniform electric field. 

The number of accessible states in arbitrary dimensions is given by
\begin{eqnarray}
%
N=\frac{V_{d}}{\left(2\pi\right)^{d}}\int{d}^{d}k,
\end{eqnarray}
where $V_{d}$ is the volume of the reservoir in the $d-$hypersphere and $d\Omega_{d-1}$ is the solid-angle element.  

To evaluate the infinitesimal element $d^{d}k$, we will employ the hyperspherical coordinates, which give us $d^{d}k=d\Omega_{d-1}dkk^{d-1}$. The differential solid angle $d\Omega_{d-1}$, in turn, takes the form
\begin{eqnarray}
d\Omega_{d-1}=\prod_{n=1}^{d-1}d\theta_{n}sin^{n}\theta_{n}.   
\end{eqnarray}

A point in the hypersphere can be specified by $(d-1)-$angles $\left(\theta_{1},\theta_{2},...,\theta_{d-2},\varphi\right)$, where $\{\theta_{k}\}_{k=1,...,d-2}\in[0,\pi]$ are the polar angles and $\varphi\in[0,2\pi]$ is the azimuthal angle. 

To obtain the spectral frequency distribution, one needs to transform the infinitesimal element in the phase space to the frequency $\nu-$space, which involves both the phase $v_{p}$ and group $v_{g}$ velocities, as usual. On the other hand, from relations (\ref{frequency1}) and (\ref{groupvelocity1}), one promptly notes that the obtained integrand is very hard to solve. To simplify our task, we then consider the weak field approximation for arbitrary angles, which is satisfied by the condition $c_{1}+d_{1}(\mathbf{\hat{k}}\cdot\mathbf{E})^{2}\gg0$. Under this assumption, both group and phase modulus velocities are equal and given by
\begin{eqnarray}
v_{g}=v_{p}\approx{c}\left(1-\frac{d_{1}E^{2}sin^{2}\theta_{d-2}}{2(c_{1}+d_{1}E^{2})}\right).
\end{eqnarray}

Therefore, taking into account the group and phase velocities in the weak field regime, we can write the number of available states $N$ as
\begin{eqnarray}\label{Nstates}
%
N=\frac{V_{d}}{c^{d}}\int{d\Omega_{d-1}}\int_{0}^{\infty}d\nu\nu^{d-1}\Delta\Lambda\left(E,\theta_{d-2}\right),
\end{eqnarray}
where
\begin{eqnarray}\label{factor}
\Delta\Lambda\left(E,\theta_{d-2}\right)\approx\left(d-1\right)+\epsilon{sin^{2}\theta_{d-2}},
\end{eqnarray}
and 
\begin{eqnarray}
\epsilon=\frac{3d_{1}E^{2}}{2(c_{1}+d_{1}E^{2})}. 
\end{eqnarray}

The factor $\Delta\Lambda$ depends on the strength of the background uniform electric field, which is encoded in the parameter $\epsilon$, and the angle $\theta_{d-2}$ between the wave vector $\mathbf{k}$ and the electric field 
$\mathbf{E}$. 

With regards to the partition function $\mathcal{Z}$, following the standard procedure of statistical mechanics for the distribution of bosonic particles with a chemical potential equal to zero, we have
\begin{eqnarray}\label{logZ}
log\mathcal{Z}=-\frac{V_{d}}{c^{d}}\int{d\Omega_{d-1}}\int_{0}^{\infty}d\nu{\nu^{d-1}}\Delta\Lambda{log\left(1-e^{-\beta{h\nu}}\right)},\nonumber\\   
\end{eqnarray}
where $\beta=1/k_{B}T$, $h$ is the Planck constant and $k_{B}$ is the Boltzmann constant.

Now that we have the logarithmic function of the partition function, we can then obtain the corresponding thermodynamic quantities associated with the photon gas, as well as the spectral energy density and the Stefan-Boltzmann law in the present scenario. 

\subsection{The spectral radiance distribution}\label{spectralradiance}

From the partition function (\ref{logZ}), it is straightforward to obtain the spectral energy density $u$, per unit volume, in thermal equilibrium at temperature $T$, which is given by
\begin{eqnarray}\label{spectralenergydensity}
u(\nu,T)=2\left(\frac{\sqrt{\pi}}{c}\right)^{d}\frac{\left(d-1\right)}{\Gamma(d/2)}\left(\frac{h\nu^{d}}{e^{\beta{h\nu}}-1}\right)\left[1+\frac{\epsilon}{d}\right].
\end{eqnarray}

\begin{center}
\begin{figure}[htb]
\begin{minipage}{0.5\textwidth}
\begin{tikzpicture}
  \node (img)  {\includegraphics[scale=0.65]{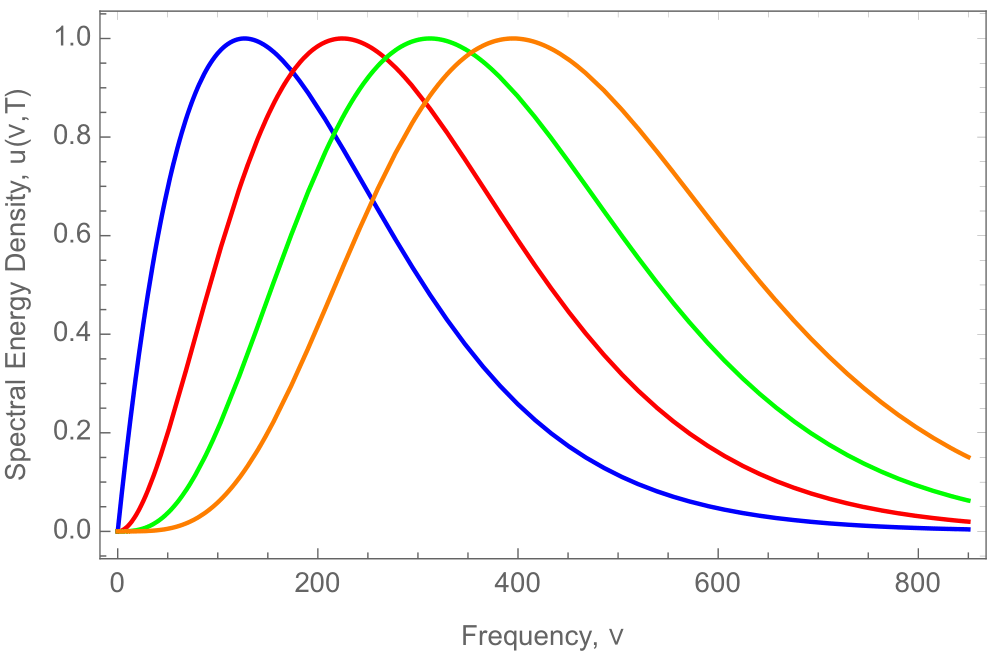}};
 \end{tikzpicture}
\end{minipage}%
\caption{Graph of the normalized spectral energy density distribution for $d=2$ (blue line), $d=3$ (red line), $d=4$ (green line) and $d=5$ (orange line). Furthermore, we have assumed a temperature of $T=0.5keV$ and $\epsilon=0.01$. Here, we adopted $c=\hbar=k_{B}=1$.}\label{FigureDimensions}
\end{figure}
\end{center}

A quick glance at the energy density (\ref{spectralenergydensity}) clearly shows us that the contributions from the nonlinearities are encoded in the $\epsilon$ parameter. In the limit $\epsilon=0$, i.e., whenever the electric field modulus $E\rightarrow0$, or, equivalently, the parameter $d_{1}\rightarrow{0}$, the internal energy density $u\left(\nu,T\right)$ reduces to the Planck distribution in large extra dimensions\cite{Cardoso:2005cd}. The Planck formula is recovered when $d=3$ and $\epsilon=0$, as expected. Furthermore, the dimension dependence $d$ in $\epsilon/d$ has a geometric origin since it arrives from the angular integration of the factor $\Delta\Lambda\left(E,\theta_{d-2}\right)$ in (\ref{logZ}). In Fig. \ref{FigureDimensions}, we have plotted the normalized spectral energy density for $d=2,3,4$ and $d=5$. Note that the frequency in which the distribution reaches a maximum depends on the dimensionality. This feature, known as Wien's law, will be discussed in the next section. 

\begin{center}
\begin{figure}[htb]
\begin{minipage}{0.5\textwidth}
\begin{tikzpicture}
  \node (img)  {\includegraphics[scale=0.65]{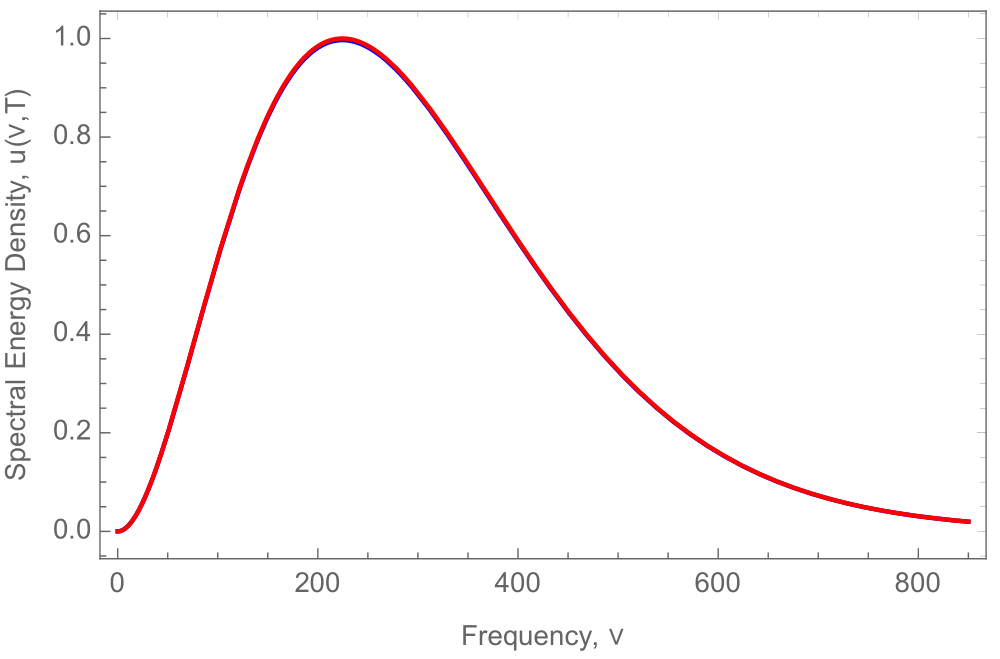}};
 \end{tikzpicture}
\end{minipage}%
\caption{Graph of the normalized spectral energy density distribution for the Planck distribution (red line) and for the case in which $\epsilon=0.001$ (blue line). In addition, we have assumed a temperature of $T=0.5keV$.}\label{Figure-spectral-1}
\end{figure}
\end{center}

Let us now consider a particular dimension and neglect for a moment the role of dimensionality to see in which way the nonlinearity affects the frequency distribution. In such case, taking the $(3+1)-$dimension into account, we compare the Planck distribution with our model, assuming $\epsilon=0.01$ (see figure \ref{Figure-spectral-1}).

\begin{center}
\begin{figure}[htb]
\begin{minipage}{0.5\textwidth}
\begin{tikzpicture}
  \node (img)  {\includegraphics[scale=0.65]{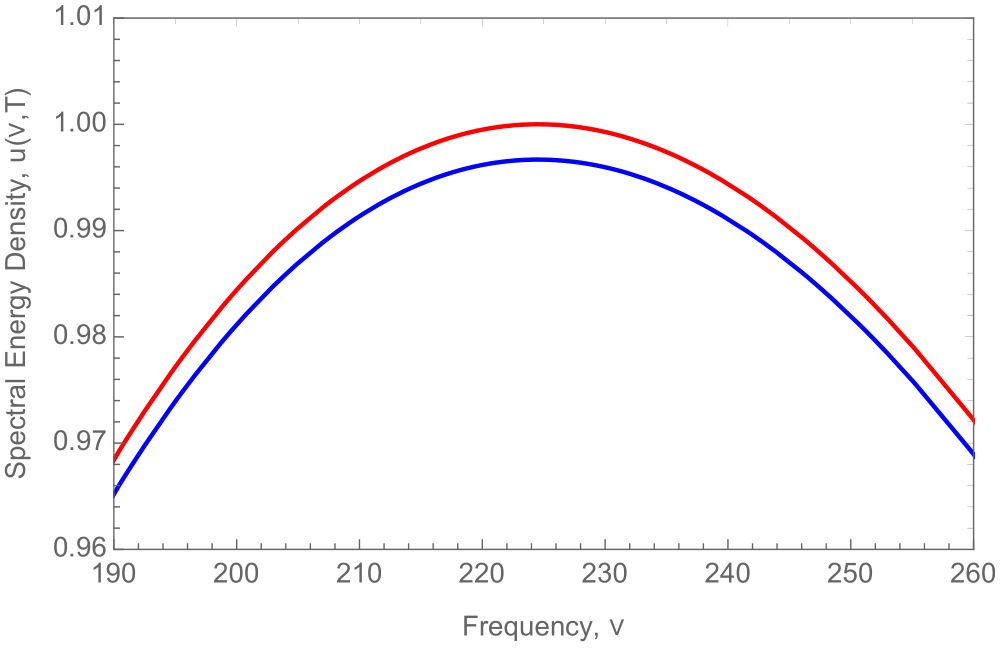}};
 \end{tikzpicture}
\end{minipage}%
\caption{Graph of the normalized spectral energy density distribution for the Planck distribution (red line) and for the case in which $\epsilon=0.001$ (blue line). Furthermore, we have assumed a temperature of $T=0.5keV$.}\label{Figure-spectral-2}
\end{figure}
\end{center}

From Fig. \ref{Figure-spectral-1}, one promptly notes that both curves are very close to each other in the scale under consideration. Therefore, performing a zoom near the pick of the curve (see Figure \ref{Figure-spectral-2}), it is possible to see an increase in the intensity of the distribution for the nonlinear case. We stress the fact that the electromagnetic nonlinearity does not change the point at which the frequency distribution reaches a maximum\cite{Soares:2023gfe}. The small difference in the intensity can be understood by analyzing the density of states. In this case, the generalized density of states $g\left(\nu\right)$ takes the following form:
\begin{eqnarray}\label{generalizeddensity}
g\left(\nu\right)=2\left(d-1\right)\left(\frac{\sqrt{\pi}}{c}\right)^{d}\frac{\nu^{d-1}}{\Gamma(d/2)}\left[1+\frac{\epsilon}{d}\right].    
\end{eqnarray}

The generalized density of states (\ref{generalizeddensity}) leads to more accessible states, which increase the number of photons for each mode. Note that this behavior depends on both the nonlinearity and the dimensionality. In other words, the nonlinear effects of the electromagnetic field induce more accessible states and higher spatial dimensions.

Next, integrating (\ref{spectralenergydensity}) over all the frequencies, the total energy density obtained is 
\begin{eqnarray}\label{totalenergydensity}
u\left(T\right)=a_{d}^{eff}T^{d+1},
\end{eqnarray}
where
\begin{eqnarray}
a_{d}^{eff}&=&\left(\frac{2}{hc}\right)^{d}\left(\sqrt{\pi}\right)^{d-1}d\left(d-1\right)k_{B}^{d+1}\times\nonumber\\
&&\Gamma\left(\frac{d+1}{2}\right)\zeta(d+1)\left(1+\frac{\epsilon}{d}\right),
\end{eqnarray}
being an effective coefficient that retains the nonlinear modifications. 

With regards to the angular dependence, the energy density contribution for each solid-angle element is given by
\begin{eqnarray}
u\left(T,\Omega_{d-1}\right)d\Omega_{d-1}&=&\left(\frac{1}{ch}\right)^{d}\left[\left(d-1\right)+\epsilon{sin^{2}\theta_{d-2}}\right]\times\nonumber\\
&&k_{B}^{d+1}T^{d+1}{d}\Omega_{d-1}.
\end{eqnarray}

Here, we would like to stress the fact that there is the appearance of an angular dependence on the power spectrum, providing an anisotropic term to the frequency distribution. 


\subsection{The generalized Stefan-Boltzmann law}\label{sectiongeneralizedSB}

The radiance is achieved by performing the integration of the spectral radiance over the hemisphere $(d-1)$-dimensional 
\begin{eqnarray}\label{radianceintegral}
R\left(T\right)=\int{d\Omega_{d-1}cos\theta_{d-2}}\int_{0}^{\infty}d\nu{B}\left(\nu,\theta_{d-2},T\right),    
\end{eqnarray}
where the spectral radiance is given by
\begin{eqnarray}
{B}\left(\nu,\theta_{d-2},T\right)=\left(\frac{\nu}{c}\right)^{d-1}\left(\frac{h\nu}{e^{\beta{h\nu}}-1}\right)\Delta\Lambda\left(B,\theta\right).    
\end{eqnarray}

Therefore, performing the integral, one promptly finds
\begin{eqnarray}\label{generalizedSB}
R\left(T\right)=\sigma_{d}^{eff}T^{d+1},    
\end{eqnarray}
where
\begin{eqnarray}\label{generalizedSBconstant}
\sigma_{d}^{eff}&=&\left(\frac{2}{c}\right)^{d-1}\left(\sqrt{\pi}\right)^{d-2}\frac{k_{B}^{d+1}}{h^{d}}d\left(d-1\right)\times\nonumber\\
&&\Gamma\left(\frac{d}{2}\right)\zeta\left(d+1\right)\left[1+\frac{\epsilon}{d+1}\right],    
\end{eqnarray}
which reduces to the previous cases when $\epsilon=0$\cite{Cardoso:2005cd}, where $\xi$ is the Riemann zeta function.

Relation (\ref{generalizedSB}) is the generalized Stefan-Boltzmann law in large infinitely extra dimensions in the presence of a background electric field, with (\ref{generalizedSBconstant}) being the generalized Stefan-Boltzmann constant. Furthermore, from relation (\ref{generalizedSB}), one concludes that the radiance is a monotonic increasing function with the number of spatial dimensions.

\subsection{The long wavelength regime and the Wien's displacement law}\label{longwave}

At low frequencies, the frequency distribution (\ref{spectralenergydensity}) assumes the form
\begin{eqnarray}
u(T)=2\nu^{d-1}\left(\frac{\sqrt{\pi}}{c}\right)^{d}\frac{\left(d-1\right)}{\Gamma(d/2)}\left[1+\frac{\epsilon}{d}\right]\left(k_{B}T\right).
\end{eqnarray}

The above relation shows us that the nonlinearities induced by the background electric field in extra dimensions modify the Rayleigh-Jeans law. On the other hand, the dimensionality does not influence the temperature, which has the same behavior as in $D=3+1$. 

With regards to the Wien's displacement law, we define $x=h\nu_{max}/k_{B}T$ in (\ref{spectralenergydensity}) and perform $du(x)/dx=0$, which gives us
\begin{eqnarray}
e^{-x}=1-\frac{x}{D},   
\end{eqnarray}
where $D$ is the spacetime dimension.

The solution of the above equation is given by 
\begin{eqnarray}
x=D+W\left(-De^{-D}\right),    
\end{eqnarray}
where the Lambert function $W$ is defined as
\begin{eqnarray}
W\left(z\right)+e^{W\left(z\right)}=z.
\end{eqnarray}

Therefore, Wien's law takes the form $\nu_{max}/T=\alpha$, where $\alpha$ is a constant\cite{Cardoso:2005cd}. It is important to point out that the electromagnetic nonlinearities do not alter Wien's law. However, the constant $\alpha$ depends on the dimensionality of the spacetime, which increases as the number of dimensions increases. An inspection of Fig. \ref{FigureDimensions} demonstrates the expected behavior of Wien's law in higher dimensions, which depends on the value of $\alpha$.


\section{Thermodynamics properties of the photon gas}
\label{thermoproperties}

Features related to the nonlinearities of the abelian spin-1 sector in large extra dimensions can be further investigated by evaluating the thermodynamic variables of the photon gas. In this sense, the straightforward way to obtain these quantities at thermal equilibrium is by obtaining the Helmholtz free energy $F$ in this situation, namely, 
\begin{eqnarray}\label{freeenergy}
F&=&-\frac{V_{d}}{d}\left(\frac{2}{hc}\right)^{d}\left(\sqrt{\pi}\right)^{d-1}d\left(d-1\right)k_{B}^{d+1}\times\nonumber\\
&&\Gamma\left(\frac{d+1}{2}\right)\zeta(d+1)\left(1+\frac{\epsilon}{d}\right)T^{d+1}.
\end{eqnarray}

From the free energy (\ref{freeenergy}), the pressure $p$, the energy $u$, the entropy $s$ and the heat capacity $c_{V}$ at constant volume densities are, respectively, given by
\begin{eqnarray}
p=-\frac{\partial{F}}{\partial{V}}, \quad s=-\frac{\partial{F}}{\partial{T}}, \quad c_{V}=\frac{\partial{E}}{\partial{T}},    
\end{eqnarray}
which give us
\begin{eqnarray}\label{pressure}
p=\frac{1}{d}a^{eff}_{d}T^{d+1}, \quad 
s=\frac{\left(d+1\right)}{d}a^{eff}_{d}T^{d},
\end{eqnarray}
and
\begin{eqnarray}\label{heat}
c_{V}=\left(d+1\right)a^{eff}_{d}T^{d}.
\end{eqnarray}

The thermodynamic quantities in (\ref{pressure}) and (\ref{heat}) show us that the electromagnetic wave propagation in a background electric field induces deviations of the free energy and the corresponding derived thermodynamic equilibrium quantities. In addition, taking into account Eqs. (\ref{totalenergydensity}) and (\ref{pressure}), the equation of state assumes the form
\begin{eqnarray}
p=\frac{u(T)}{d},   
\end{eqnarray}
where $d$ is associated with the spatial dimension in $M^{d+1}$.\\
\section{Further remarks}\label{further}

Let us now discuss our results. Some previous works have already considered the question of the electromagnetic radiation in thermal equilibrium in large extra dimensions\cite{Al-Jaber2003,Cardoso:2005cd,Alnes:2005ed}. 
In these works, there appears a 
generalization of the Stefan-Boltzmann law with a dependence on the spatial dimension, i.e., $R\sim{T^{d+1}}$. In our framework, in turn, we have extended these results by introducing the nonlinear behavior of the electromagnetic field. Although there exists an influence of the nonlinearity, which is encoded in the parameter $\epsilon$, the temperature dependence in the radiance presents the same behavior as the mentioned works. It can be easily understood by the fact that the temperature dependence in the Stefan-Boltzmann law, for instance, emerges when one performs the integral over the frequencies, which goes to $\nu^{d}$, as one can see from relation (\ref{radianceintegral}). The same argument is valid for the energy density and the thermodynamics quantities as well. In this sense, we can conclude that there exists a universal behavior of the Stefan-Boltzmann law in extra dimensions for electromagnetic theories that hold a linear dependence between the wave frequency $w$ and the momentum $\mathbf{k}$, which goes to $R\sim{T^{d+1}}$. Theories in which there appears a distinct relation between the frequency modes and the wave vector, such as $w\sim{k^{n}}$ with $n\ne1$, for instance, would give a different pattern to the Stefan-Boltzman law.

In Ramos\cite{Ramos:2014mda}, a more physically appealing model was studied. In their work, the authors have analyzed the influence of compact extra dimensions in the spectrum energy distribution. In the low-temperature regime, the blackbody radiation behaves as usual, while at temperatures above the inverse to the size of the compact dimensions, small deviations appear in the spectrum distribution as well as in the energy density, which depends on the dimensionality of the spacetime. Here, the role of the extra dimension emerges in the regime of high temperatures. On the other hand, in this regime, the temperature dependence is the same as in non-compact extra dimensions.

Regarding the birefringence in wave propagation, we have shown that the nonlinearity, when considering both external magnetic and electric fields, percolates through one single mode. It is because we have considered nonlinear theories that depend exclusively on the invariant $\mathcal{F}$. In the $(3+1)-$case, for instance, theories such as Euler-Heisenberg and Born-Infeld take into account invariants that depend also on the $\mathcal{G}\left(=-(1/4)F^{\mu\nu}\Tilde{F}_{\mu\nu}=\mathbf{E}\cdot\mathbf{B}\right)$. In such a case, there is the emergence of two distinct modes that are field-dependent. The extension of the invariant $\mathcal{G}$ for arbitrary dimensions, in turn, is not clearly-cut since it depends on the dimensionality of the spacetime, i.e., the electromagnetic dual tensor $\tilde{F}^{\mu\nu}$ is a function of the Levi-Civita symbol, which depends on the spacetime dimension. 

Concerning the role of spin-statistics and dimensionality of space, a few comments are in order. The spin-statistics theorem states that integer spin particles follow the Bose-Einstein distribution, while half-integer spin particles obey the Fermi-Dirac distribution. This theorem emerges from the Relativistic Quantum Field Theory in four spacetime dimensions. For spatial dimensions different from $d=3$, this connection is not guaranteed. The role of spin and statistics has been investigated by some authors. Weinberg\cite{Weinberg:1984vb}, for instance, studied the connection between spin and statistics for massive particles and showed that the relation between bosons and fermions is valid for all spacetime dimensions. Boya and Sudarshan\cite{Boya:2007pc}, in turn, have investigated if the symmetry of the bilinear scalar product in the lagrangian under rotations in $d=3$ can be extended for quantum fields in arbitrary dimensions. They found that the standard connections between bosons and fermions are valid for specific dimensions. Here, we have adopted the point of view that the Bose-Einstein statistics remain valid independent of the spacetime dimension, where each mode has an average energy $\Bar{E}$ given by the Planck's formula, namely, 
\begin{eqnarray}
\Bar{E}=\frac{h\nu}{e^{\beta{h}\nu}-1}.    
\end{eqnarray}


\section{Conclusions}\label{conclusions}

In this work, we have studied the nonlinear electromagnetic wave propagation in large extra spatial dimensions. Specifically, we have obtained the wave frequencies and the group velocities for both background electric and magnetic fields. Furthermore, the generalized blackbody radiation laws and the thermodynamics quantities at thermal equilibrium for a photon gas in the presence of an external constant electric field were also contemplated. 

We also would like to remark that the configuration studied in this work seems to reproduce a nonstandard physical situation. It occurs because the spatial extra dimensions are assumed to be infinite, while modern theories suggest that these spatial extra dimensions must be compactified up to a very small scale. On the other hand, theoretical models only can be ruled out by experiments, which, at this point, is not the case. In addition, it is the first time, up to our knowledge, that nonlinear theories of electrodynamics are contemplated in a general formalism that depends on extra spacetime dimensions. Indeed, it can provide us with valuable information about some features of electromagnetism in this scenario. 

To conclude, we would like to highlight that eventual deviations from the well-known blackbody radiation thermal laws in $(3+1)-$dimensions would be an indication of new physics. In this vein, a lab setup in which a photon gas is under the influence of an external electric field could unveil the existence of these extra dimensions through deviations from the Planck distribution, similar to those present in Figs. \ref{FigureDimensions}, \ref{Figure-spectral-1} and \ref{Figure-spectral-2}. Another possibility to probe extra dimensions would be related to the Bose-Einstein condensate in the presence of external electric and magnetic fields. Indeed, it was experimentally observed that a thermalization process for a two-dimensional photon gas in a dye-filled optical microcavity gives rise to a condensate of photons\cite{Klaers2010}. The influence of background electromagnetic fields as well as extra spatial dimensions during the condensation process could induce modifications at the transition temperature, for instance, which could reveal the existence of these extra dimensions. We intend to explore this possibility in the foreseeable future. We hope that the present work will stimulate further investigations into the issue.\\

\section{Acknowledgements}

This work is a part of the project INCT-FNA proc. No. 464898/2014-5. RT acknowledges financial support from the PCI program of the Brazilian agency Conselho Nacional de Desenvolvimento Científico e Tecnológico – CNPq. SBD thanks CNPq for partial financial support.



%
%
%




\bibliographystyle{apsrev4-1}
\bibliography{Bibliography.bib}

\end{document}